\documentclass[preprint,3p,times,twocolumn]{elsarticle}
\usepackage{graphics}
\usepackage{epsfig}
\usepackage{amssymb}
\usepackage{lineno}
\journal{Nuclear Instruments and Methods A}

\begin{document}

\begin{frontmatter}

\title{Direct measurements of the properties of Thick-GEM reflective photocathodes}

\author[label2]{M. Baruzzo\fnref{present-address-UD}}	
\author[label3]{C. Chatterjee}   
\author[label3]{P. Ciliberti}   
\author[label2]{S. Dalla Torre}	
\author[label3]{S.S. Dasgupta\fnref{leave-matrivani}}	
\author[label2]{B. Gobbo}		
\author[label2]{M. Gregori}		
\author[label1]{G. 
Hamar\corref{cor}}
\author[label2]{S. Levorato}		
\author[label2]{G. Menon}	
\author[label2]{C. A. Santos}	
\author[label2]{F. Tessarotto}	
\author[label2]{Triloki\fnref{ictp}}
\author[label1]{D. Varga}		
\author[label2]{Y. X. Zhao}

\address[label2]{INFN Sezione di Trieste, Trieste, Italy}
\address[label3]{University of Trieste and INFN Sezione di Trieste, Trieste, Italy}
\address[label1]{Wigner Research Centre for Physics, Budapest, Hungary}
\address[label4]{Physics Department, University of Aveiro, Aveiro, Portugal}

\cortext[cor]{Corresponding author. E-mail address: gergo.hamar@ts.infn.it (G.Hamar)}
\fntext[present-address-UD]{present address: University of Udine, Udine, Italy and INFN, Sezione di Trieste, Trieste, Italy}
\fntext[leave-matrivani]{on leave  from Matrivani Institute of 
Experimental Research and Education, Kolkata, India}	
\fntext[ictp]{also at Abdus Salam ICTP, Trieste,Italy}

\begin{abstract}
In the context of the development of novel Thick GEM based 
detectors of single photons, the high resolution 
optical system, nicknamed Leopard, providing 
a detailed surface scanning of the Thick GEM 
electron multipliers, has been used for a set 
of systematic measurements of key Thick GEM 
properties. These results are reported and discussed. 
They confirm by direct observation Thick GEM properties 
previously inferred by indirect measurements 
and answer to relevant questions related to 
the use of Thick GEMs as photocathode substrates 
in novel gaseous photon detectors.
\end{abstract}


\end{frontmatter}

\tableofcontents


\section{Introduction}
\label{introduction}
The first generation of gaseous photon detectors with 
solid state photocathode are the MultiWire Proportional
Chambers (MWPC) coupled to a CsI photocathode\cite{rd26}. 
Novel gaseous photon detectors must represent a 
progress in the field and thus they must match 
two basic requirements, namely reduced rates of 
photon backflow and of Ion BackFlow (IBF) to 
the photocathode.  These reduced rates help to overcome 
the photocathode ageing, make possible high 
gain operation and intrinsically fast detector operation, 
Essentially, an achievement of detectors with high rate capabilities. 
Recently developed Micro Pattern Gaseous Detectors (MPGD)
look promising concerning both of these issues.
Gas Electron Multiplier (GEM)\cite{gem} and GEM-derived 
multipliers, as THick GEMs (THGEM)\cite{thgem}, 
are intrinsic fast devices. In fact, the signal 
is mainly generated by the electron motion. 
In MICROMEGAS (MM)\cite{mm}, in spite of the parallel 
plate structure, the signal development is 
fast thanks to the extremely thin cathode-anode gap. 
Concerning the suppression of the photon and ion backflow, in multilayer 
GEM and THGEM structures, no photon feedback is 
present, while a good fraction of the ions is 
trapped in the intermediate layers 
and do not reach the photocathode; in particular 
dedicated studies of the IBF rates have been performed 
due to the interest not only for photon detection\cite{thgem-ibf}, 
but also for the use of GEMs as TPC read-out 
elements\cite{alice-tpc}.  
In MMs, the ions created in the multiplication process are 
naturally trapped in the multiplication gap thanks to the 
large unbalance of the electric field above and below 
the micromesh which defines the amplification region.
\par 
The concept of MPGD-based photon detectors had 
a first application in the threshold Cherenkov counter 
Hadron Blind Detector (HBD)\cite{hbd} of the Phenix experiment: 
the photon detectors are triple GEM counters where the 
first GEM foil,  coated with a CsI film, is used as reflective 
photocathode; the detectors are operated at a low gain 
level (4000), not adequate for efficient single photon detection. 
Thanks to the observed high gain, THGEMs have been
proposed for novel gaseous detectors of single photons\cite{thgem-amos} 
and relevant R\&D studies have been 
performed\cite{thgem-ibf, thgem-alice,thgem-noi, gain-time-evolution-paper, noi_2010_qe}.
THGEMs can be used in multilayer arrangements or coupled to 
a MM multiplication stage, as it is the 
case for the upgrade of the COMPASS RICH.
In all  these architectures the photocathode substrate is a THGEM plate.
In this context, the opportunity of mapping the THGEM 
response to single UV photons offers a handle of great 
relevance for understanding the behavior of THGEM photocathodes. 
These studies are made possible by the  system for the high 
resolution surface scanning of THGEMs by single photo-electron 
detection, nicked named Leopard, 
recently introduced\cite{leopard} (Sec.\ref{leopard}). 
A number of central questions related to the development 
of THGEM based novel photon detectors have been answered 
thanks to the Leopard capability to provide detailed gain and efficiency maps. 
\par
This article is dedicated to a first set of systematic studies 
of THGEM multipliers, which are introduced in Sec.~\ref{sec:thgem},
while the THGEMs used as photocathodes for the upgrade of the 
COMPASS RICH are described in  Sec.~\ref{sec:thgem-pc}. 
The measurements have been performed by the Leopard system, 
described together with the 
overall experimental setup in Sec.~\ref{setup}; the measurement 
procedures are given in Sec.~\ref{procedures}, while
the results about gain and photoelectron extraction are 
presented and discussed in 
Secs.~\ref{gain} and \ref{extraction}, respectively.
These studies are performed converting the 
light in the gold coating of the THGEM PCB surface. 
Dedicated measurements reported in Sec.~\ref{csi-gold}  
allow to extend the results concerning photoelectron extraction from gold 
to photoelectron extraction from CsI.
Conclusions are presented in Sec.~\ref{conclusions}.

\section{The THGEM electron multiplier}
\label{sec:thgem}

\begin{figure}
    \begin{center}
    \includegraphics[width=.45\textwidth]{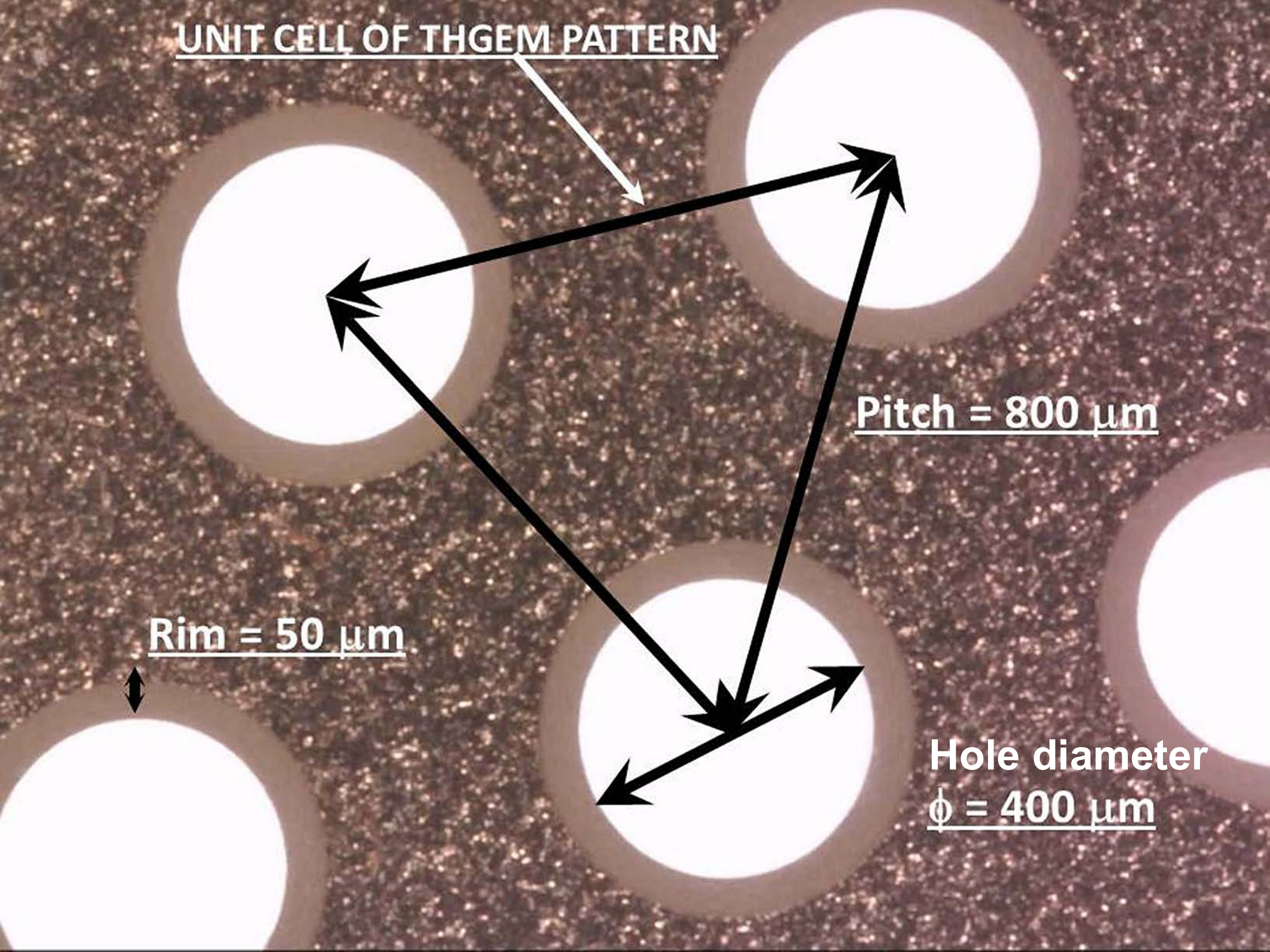}
    \caption{\emph{Detail of a THGEM PCB (picture).}
    \label{fig:thgem_picture}}
    \end{center}
    \end{figure}

THGEMs, introduced in parallel by several groups\cite{thgem}, 
are electron multipliers derived from the GEM design, 
scaling the geometrical parameters and changing the 
production technology. The Cu-coated polyimide foil of 
the GEM multipliers is replaced by standard PCBs and the 
holes are produced by drilling. The conical shape of the 
GEM holes that forms uncoated polyimide rings around the 
holes themselves are replaced by a clearance ring, the rim, 
surrounding the hole and obtained by Cu etching. The hole 
arrangement is similar to the one adopted for the GEMs; 
the circular hole centres are distributed according to a 
repetitive pattern: the basic cell is an equilateral triangle. 
Typical values of the geometrical parameters are 
PCB thickness of 0.2-1 mm, hole diameter ranging between 0.2 and 1 mm, 
hole pitch of 0.5-1.2 mm and rim width between 0 and 0.1 mm (Fig.\ref{fig:thgem_picture}). 
The early phase of the THGEM characterisation 
was largely contributed by the Weizmann 
group led by A. Breskin\cite{thgem-amos}. 
In this context, large gains along with good rate 
capabilities have been reported for  
single or double THGEM layers.
\par
THGEMs can be produced in large series and large size 
with standard PCB technology, instead of large number 
of holes present, few millions per square meter.
THGEMs have intrinsic mechanical stiffness, 
and they are robust against damages produced 
by electrical discharges. Due to the technology used, 
the material budget of THGEM based detectors is not 
particularly reduced, and due to the enlarged geometrical 
parameters, they cannot offer space resolution as fine as 
GEM based detectors. Thanks to the reduced gaps between the 
multiplication stages, THGEM based detectors can be successfully 
used in magnetic field. These features, shortly mentioned 
above, match very well the requirements of specific applications 
in fundamental research, where the large gain, the robustness, 
the production technique and the mechanical characteristics 
are advantages, while the material budget and the space 
resolution aspects do not represent a limit. THGEMs are 
considered for the single photon detection in Cherenkov 
imaging applications, as active elements in hadron 
sampling calorimetry\cite{h-cal}, for muon tracking\cite{mu-tracking} and for 
the read-out of noble liquid detectors\cite{lar}.

\section{The THGEM-based photocathodes of the photon 
detectors for the upgrade of COMPASS RICH-1}
\label{sec:thgem-pc}

\begin{figure}
    \begin{center}
    \includegraphics[width=.5\textwidth]{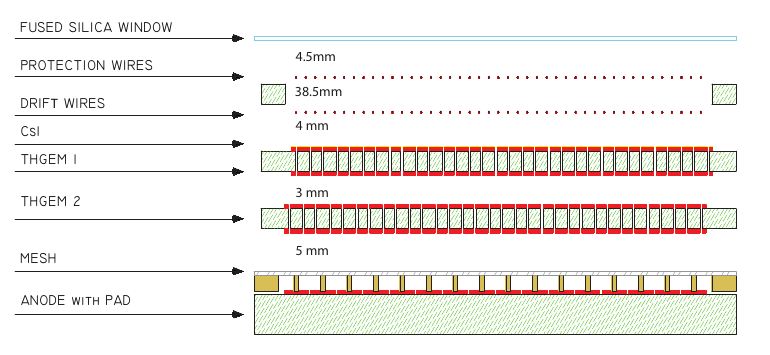}
    \caption{\emph{Scheme (not to scale) of the hybrid detector architecture 
    for the COMPASS RICH upgrade. It includes two staggered THGEM layers 
    (thickness: 0.4mm; hole diameter: 0.4mm; pitch: 0.8mm), 
    and a MM (128~${\mu}$m gap, multi-pad anode).}
    \label{fig:hybrid}}
    \end{center}
    \end{figure}

Novel MPGD-based photon counters with a hybrid architecture 
formed by two layer of THGEMs and a layer of MM  
(Fig.\ref{fig:hybrid}) have being developed and built\cite{hybrid-upgrade}
for the upgrade of COMPASS\cite{compass} RICH-1\cite{rich1},
implemented in 2016.
The first THGEM acts as photocathode substrate and its 
upper surface is coated with a thin (60~nm) CsI film.
CsI has non negligible Quantum Efficiency (QE) 
in the far UV domain, at photon wavelength shorter than 210~nm. 
The effective QE in gas atmosphere has been studied 
by several authors\cite{csi-gas} and proven by the RD26 development\cite{rd26}
and the use of MWPCs with CsI photocathodes in several
experiments\cite{csi-experiments, rich1}: adequate gas atmosphere 
is required as well as the presence at the photocathode surface of
an electric field exceeding 500~V/cm.
CsI is chosen because, among the typically used 
solid state photoconverting materials,
it is definitively the most robust one, thanks to a work function higher than
the other frequently used photon converters, in particular 
the ones with QE in the visible-light range. 
CsI also exhibits a relative chemical
robustness against oxygen and water, as compared to other photoconverters,
especially visible-sensitive ones: its QE is preserved in case of
short exposure to air, namely to oxygen and water vapour, while, on a term
basis of years, it can tolerate the exposure to atmospheres with oxygen and
water vapour contamination at the level of a few ppm\cite{csi-long-term}. 
\par
It is relevant to notice that the configuration adopted 
for the upgrade of COMPASS RICH-1 is the reflective photocathode one, 
which is preferred compare to the architectures 
with semitransparent photocathode, as it results
in a larger photoconversion rate. In fact, a semitransparent one
requires the application of a thin metallic film, 
which absorbs photons, to keep
the entrance window at a fix potential; also the probability of photoelectron
absorption is lower in a reflective photocathode than in a semitransparent
one as the conversion probability is the highest at the entrance surface of
the photoconveter. Moreover, the thickness of the coating layer is largely non
critical in the reflective configuration and this fact opens the way to the 
realization of large surface detectors.
\par
The hybrid MPGD photon detector IBF rate is at the
3\% level and it can be operated at gains higher than 10$^4$\cite{thgem-elba}.
The photoelectron extraction properties have been already studied 
by indirect measurements and are further investigated by the 
measurements reported in the present article.

\begin{figure}
    \begin{center}
    \includegraphics[width=.5\textwidth]{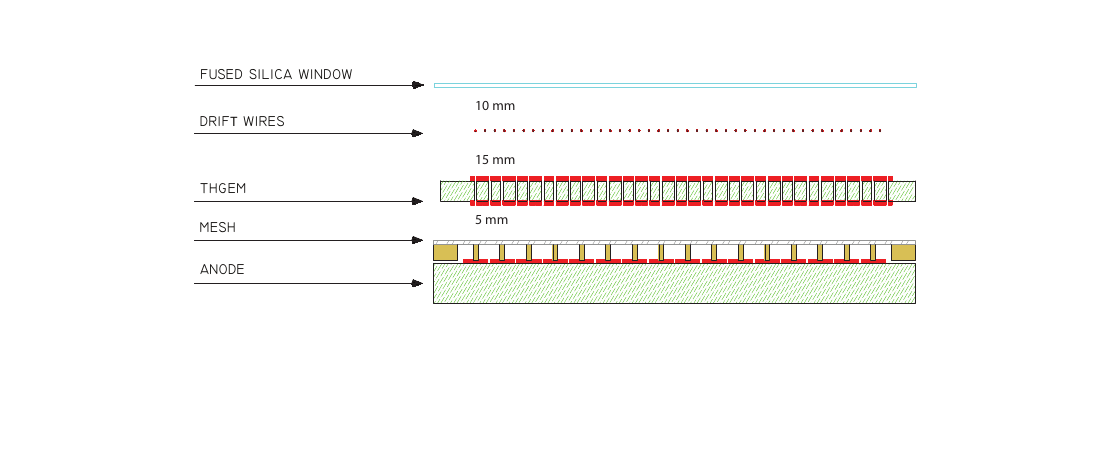}
    \caption{\emph{Scheme of the hybrid detector including a THGEM 
    used for the measurements described in this article (not to scale).}
    \label{fig:setup}}
    \end{center}
    \end{figure}

\section{Equipment and setup}
\label{setup}

\subsection{The scanning system and the setup}
\label{leopard}

The capability of the Leopard scanning system 
of performing  high resolution scanning of THGEMs 
providing independent gain and efficiency maps by the detection of 
single photoelectrons
has been proven\cite{leopard}.
The nickname "Leopard" has risen from the first images, where the photon-yield map
looks like the fur marks of the wild animal.
After the first principle tests, the system has 
been upgraded. In the following, we describe 
the system version used for the measurements reported in the present article.
\par
The Leopard system consists of a pulsed UV light 
source with a focusing optics, mounted 
onto a three-dimensional moving table, 
and a fast data acquisition and control system reading 
a single channel ADC digitizing the amplified signals from the detector
housing the THGEM under study.
The \textbf{light source} is an LED providing 245~nm 
wavelength light\footnote{Senson Technology Inc. UVTOP-240
	www.s-et.com, http://www.s-et.com/spec-sheets/240nm-with-images.pdf}.
The size of the light-spot is dictated by a dedicated pinhole.
The measurements reported in the present paper have been
performed with a 150~$\mu$m pinhole, resulting in a spot with a FWHM
size of 7~$\mu$m.
By adequate setting of  the intensity of the LED light, 
photoelectrons can be extracted from  gold- 
or copper-coated  surfaces and the single 
photoelectron mode can be established: 
for this purpose the light intensity is 
tuned in order to have no more than some 
percent of non-empty events per pulse;
for instance, the typical rate of non-empty events for the measurements 
reported in the present article is 5\%. The stability of the
intensity of the light source has been studied:
the intensity is stable  within the 2\% level over periods of 24~h.
\par
The light enters the detector via a fused silica window
and it is focused onto the top of the 
\textbf{THGEM}. The extracted photoelectrons 
are driven by the electric dipole field due to the THGEM 
voltage polarization into the nearest hole. Here they  
are amplified. The signal obtained by this single amplification stage is too 
small for effective detection and a second amplification stage
is provided by a \textbf{MM}.  A scheme of the hybrid MPGD 
used for the measurements described in this article is shown in Fig.\ref{fig:setup}. 
The gas mixture Ar:CH$_4$~=~30:70 was used for most of the 
measurements (Sec.~\ref{effective_qe}), except for the 
overnight ones when Ar+CO$_2$~=~30:70 was used because of safety considerations.
The detector has been operated at gains
in the range 1-2 10$^4$, unless explicitly mentioned.
The MM anode plane is segmented in strips 
(400~${\mu}$m pitch, 20~${\mu}$m width)
connected in parallel to the single read-out chain is use: 
the read-out area is 20 $\times$ 60~mm$^2$.
The read-out chain is formed by a preamplification stage 
followed by amplification.

\subsection{The control and data acquisition system of the scanning system}
\label{fast-leopard}

The control system of the Leopard scanning system must ensure
the adequate synchronizing among the Data AcQuisition (DAQ) system,
the data storage, the actuator system, 
with the three-dimensional movements of the light-source and the related optical 
setup, and additionally set detector parameters as well.
\par
Measurement points in steps much smaller than the typical 
dimensions of the structure under study are required to obtain detailed images.
For instance, the typical space-step, along two orthogonal axes, of the measurements 
reported in the present article is, for both axes, 100~${\mu}$m, resulting in 10$^4$
measured points per cm$^2$. 
Thousands of triggers in each measurement point are needed to get, 
point by point, a sample adequate for the extraction of the gain and the efficiency.
Typically, we have collected point to point spectra with about 1000 entries,
roughly corresponding to 20 thousand triggers per measured point. 
These requirements demand recording billions of events:
therefore, the DAQ acquisition rate is crucial.
\par
The control and high-rate DAQ functions are performed by 
a RaspberryPi~\footnote{RaspberryPi, \\ http://www.raspberrypi.org/} 
microcomputer coupled via its General-Purpose Input/Output (GPIO)
to a dedicated board.
The board receives the asynchronous trigger whose delayed rising edge serves as 
selection time for a 12-bit ADC (LTC1415~\footnote{Linear Technology, LTC1415, \\
	http://cds.linear.com/docs/en/datasheet/1415fs.pdf}) measuring the 
signal from the chamber, and the ADC is read out 
directly by the RaspberryPi via the GPIO.
Missed triggers are counted, thus allowing digital signal processing at software level.
An event rate of 130 kHz has been used for the measurements described in this article.
\par
\begin{figure}
    \begin{center}
    \includegraphics[width=.45\textwidth]{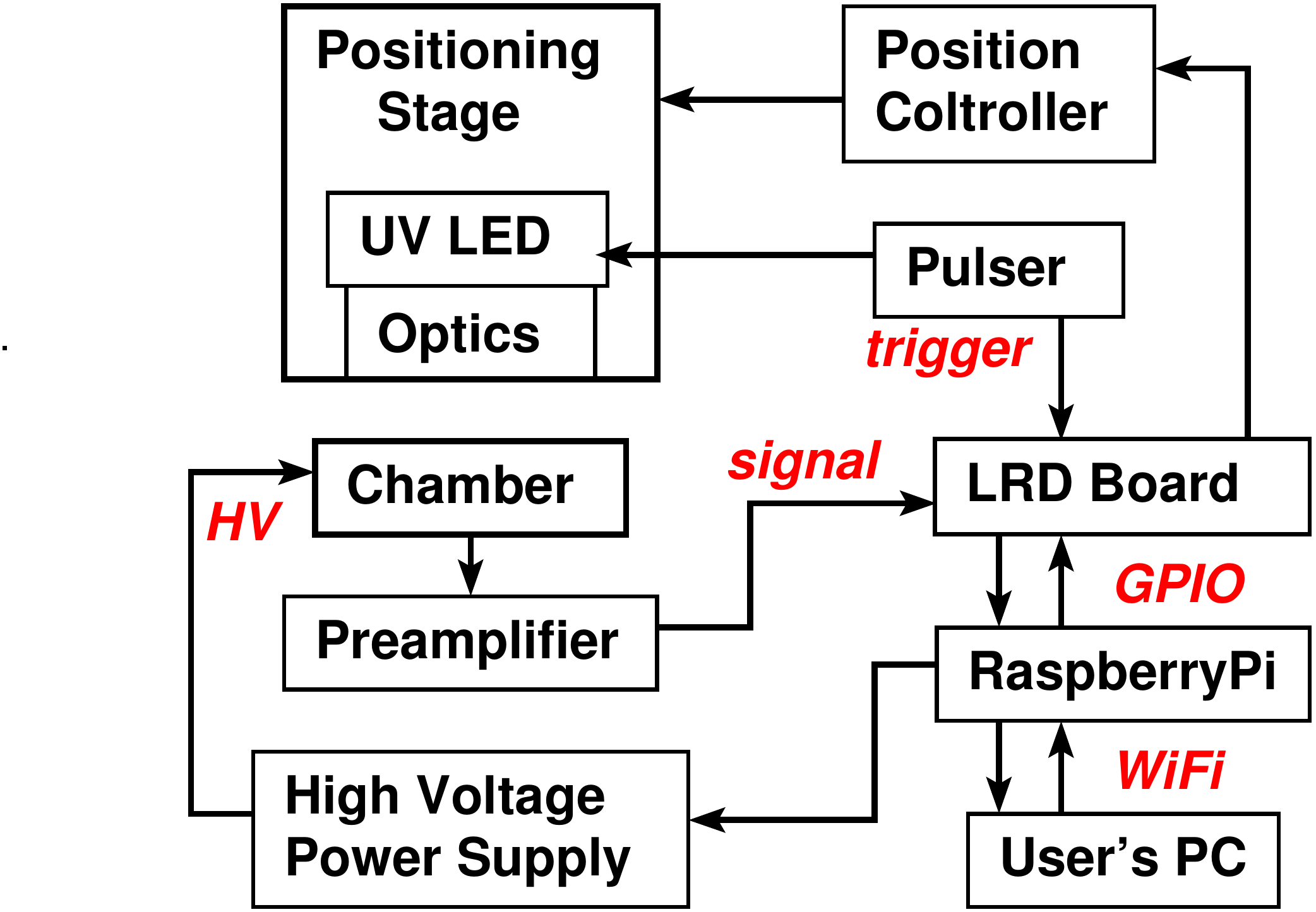}
    \caption{\emph{Block diagram of control and DAQ of the scanning system \cite{Leopard-2014-Tipp}.}
    \label{fig:LeopardSetupSchematics}}
    \end{center}
    \end{figure}

The control and DAQ software runs on the RaspberryPi 
under a standard Debian based Raspbian linux system.
The source is written in $C/C$++ to make it both fast and flexible.
The program accepts as inputs command lines  and settings files,
making convenient its remote handling.
The DAQ program is also accessible by a graphical user interface, 
which runs on a distant machine, via intranet communication,
where wireless connection is used to eliminate ground loops from Ethernet cables. A 
block diagram of control and DAQ system is presented in Fig.~\ref{fig:LeopardSetupSchematics}.

\subsection{The measured thick GEMs}
\label{thgem-list}

The measurements reported in this article refer to the THGEMs 
listed in Table~\ref{tab:thgems}, where their geometrical parameters 
are reported.

\begin{table}
\caption{\emph{List of the THGEMs used for the measurements reported 
               in the present article; 
               the geometrical parameters of the devices are provided.
               }
               \label{tab:thgems}
               }
\begin{center}
\begin{tabular}{|l||c|c|c|c|}
\hline  
	THGEM	& Hole	& Pitch	& Thickness	& Rim	\\
	        & diamter &     &           &       \\
	Name	& [$\mu$m]	& [$\mu$m]	& [$\mu$m]	& [$\mu$m] \\
	\hline
	M1-III	& 400		& 800		& 400		& 0	\\
	DESTRO-I	& 400		& 800		& 400		& 5	\\
	C3HR-II	& 400		& 800		& 400		& 50	\\
	M2.4-G	& 400		& 800		& 600		& 0	\\
	M2.1-II	& 300		& 800		& 400		& 0	\\
	\hline
\end{tabular}
\vspace{0.4cm}
\end{center}
\end{table}

\section{Measurement procedures}
\label{procedures}


The (x,y,z) reference adopted in the following has been
chosen such that the (x,y) plane is parallel
to the THGEM surface, with the y-axis 
parallel to a row of hole centres and the z-axis orthogonal 
to the THGEM surface.

\subsection{Algorithms for gain and yield extraction}
\label{algorithms}

In case of multistage detectors the distribution of the avalanche amplitude
is mostly defined by the first amplification process. 
Operating the THGEMs at moderate gain, the expected amplitude distribution 
is exponential and this is the dominating distribution in 
our setup, as it has been crosschecked in the data.
\par
Therefore, in the following, 
we assume a simple exponential distribution; in this case the most important 
parameters can be computed with small uncertainties even with limited statistics,
as the ones we have collected in each point. 
We define the yield ($Y$) as the detected number of photoelectrons, namely
the number of hits above the threshold set at a level of five standard deviation 
of the electronics noise distribution. 
The gas gain ($G$) can be calculated 
from the average or from the median of the amplitude distribution of these hits,
when the threshold value is known.
Knowing $G$, the total number of photoelectrons ($Y_{extr}$)
can be computed by a proper extrapolation.
The simple equations we use are reported in the following,
where $Q$ is the signal amplitude, $Q_{cut}$ is the threshold value, $S(Q)$ is
the amplitude distribution, $N$ is the integral of $S(Q)$ over the whole range, 
$\Theta$ is the Heaviside function, and
$Med[a:b]F$ refers to the median 
of the distribution $F$ in the range $[a:b]$.

\begin{eqnarray}
	\label{EqGpd}
	S(Q) &=& \frac{N}{G}  \cdot e^{-Q/G} \cdot \Theta(Q) \\
	Y &=& \int_{Q_{cut}}{ S(Q)} dQ = N \cdot e^{-Q_{cut}/G} \\
 	Y_{extr} &=& \int_0{ S(Q) dQ } = N = Y \cdot e^{Q_{cut}/G} \\
	Q_{med} &=& Med [Q_{cut}:\inf](S(Q)) \\
	G &=& \frac{1}{ln2} (Q_{med}  - Q_{cut} )
	\label{EqGpd2}
	\end{eqnarray}

\par
A key ingredient to study the THGEM performance on a hole-by-hole 
basis is the assignment of the correlation between an illuminated point 
and the THGEM hole where the photoelectron drifts and initiates the avalanche process.
The adopted strategy is straight-forward: 
when a point of the THGEM surface is illuminated, the
measured $G$ is related to the nearest hole. 
As a result, the area related to a hole has a hexagonal shape;
this area is referred to as hole-area in the following.
This choice is supported by electrostatic considerations: due to the 
symmetry of the THGEM geometry the field lines originated at one of the THGEM surfaces
and ending at the other surface enter the nearest hole.
Moreover,  the two-dimensional (x,~y) images obtained with poor uniformity THGEMs,
as, for instance, the one shown in\cite{leopard}, Fig.~15, clearly indicate 
that the measured $G$-maps support our strategy.

\subsection{Image focusing}
\label{focusing}

\begin{figure}[h]
	\centering
	\includegraphics[width=.45\textwidth]{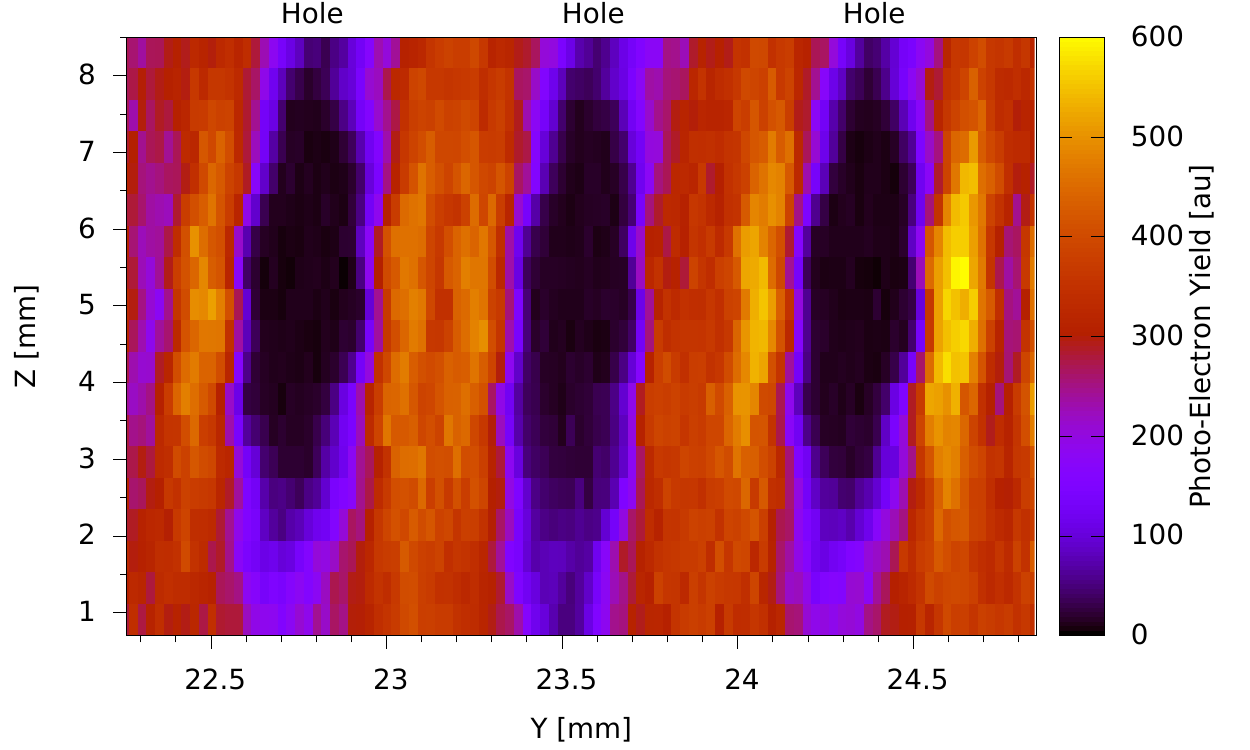}
	\caption{Image focusing by a two-dimensional scan in the (y,z) plane; 
                 for  each (y,z) point, the yield normalized to
                 a fix number of triggers is reported.}
	\label{FigFocalScan2D}
	\end{figure}

\begin{figure}[h]
	\centering
	\includegraphics[width=.45\textwidth]{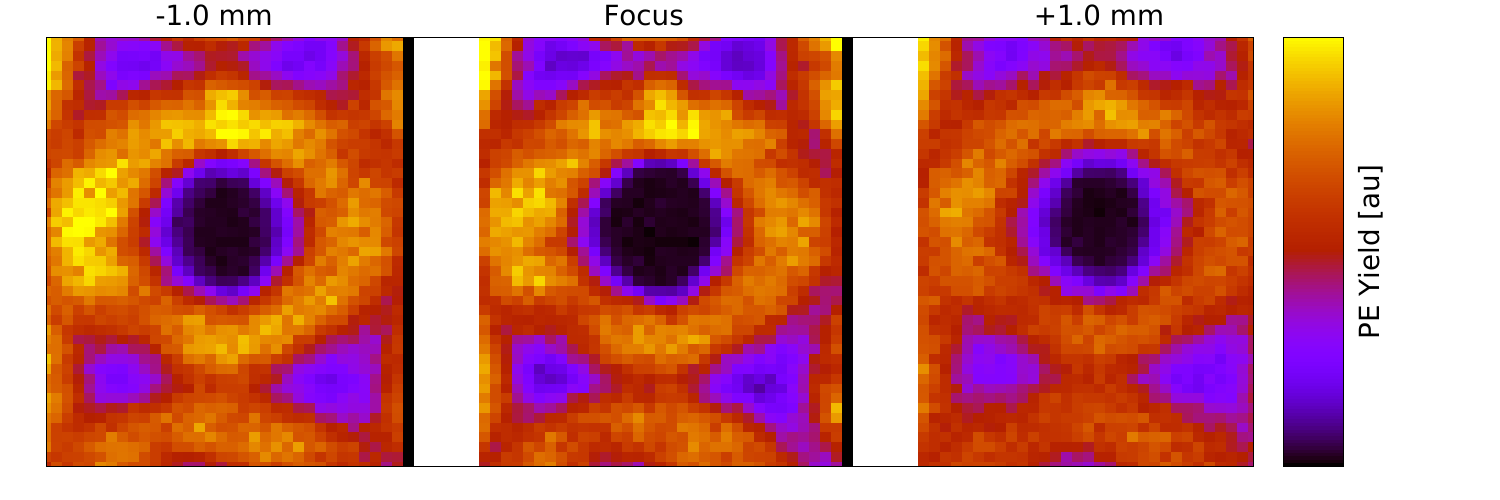}
	\caption{Yield maps at focus and slightly outside focus. It can be noticed that 
                 the change in image sharpness is limited within $\pm$1~mm around 
                 the best focal settings.}
	\label{FigFocalSearch3D}
	\end{figure}

In order to ensure high measurement resolution, whenever a THGEM plate is installed,
the focal height of the scanning table is set. For this purpose, following the procedure 
already described in \cite{leopard},
an (y,z) scans through a line of holes is performed, where the sharpest
images indicates the best focal settings.
The scan outcome for one of the settings is shown in 
Fig.~\ref{FigFocalScan2D}, where, for  each (y,z) point, the yield  normalized to
a fix number of triggers is reported. The resolution in the focal plane determination is
of the order of 1~mm. This is confirmed by the 
two-dimensional (x,y) images taken  at z-values near the focal setting
(Fig.\ref{FigFocalSearch3D}). 

\subsection{Estimation of the errors}
\label{errors}

The statistical errors on the measured quantities,
namely $G$, $Y$ and $Y_{extr}$, which are extracted from
spectra using 20.000 events with about 400-1000
photo-electron hits, are typically at the 5\%, 5\% and 10\% level, respectively.
\par
$Y$ and $Y_{extr}$ are affected by systematic errors related to the variation
of the light intensity, which is stable at the 2\% level (Sec.~\ref{setup}).
\par
The most relevant systematic error affecting $G$ is due to the effects of 
time evolution of the gain discussed in Sec.~\ref{charging-up}
and dedicated strategies are implemented to limit 
the error size,  when data dedicated to gain-maps 
are collected. The relevant systematic error for these studies is the 
one introducing point-to-point gain variations.
By repeated measurements, we estimate that the residual 
relative systematic error is around the 5\% level.
\par
The correct assignment of the measurements to the corresponding hexagonal hole area
(Sec.~\ref{algorithms}) is critical only for measurements performed at the edges of 
the hexagons in tiny perimetral corridors that have a width dictated by the
light spot size at the THGEM surface, 
namely 7$\mu$m FWHM (Sec.~\ref{setup}).
This effect is neglected in the data analysis.

\section{Charging-up aspects}
\label{charging-up}

\subsection{The time evolution of the gain in THGEM multipliers}
\label{gain-time-evolution}

Dedicated studies had been devoted to understand the  
time-evolution of the gain in THGEM multipliers\cite{gain-time-evolution-paper}.
\par
The time evolution of the THGEM gain exhibits two distinct phenomena: a fast evolution,
which is exhausted over time ranges between a few minutes up to about twenty minutes and a
long-term time evolution, which develops over days. Both effects depend 
on the amount of the open dielectric surface
present in the multiplier, which is related to the geometrical parameters.
The former is due to the charge accumulation
at the free dielectric surface present in the detector,
namely the so-called charging-up: 
the resulting charge distribution and the time
required to reach the asymptotic configuration depend on the THGEM geometry,
the applied voltage and the irradiation rate; the charge accumulated at the free dielectric surface
always reduces the electric field and thus the detector gain. 
The long-term time evolution is the more delicate aspect because of the relevance of the effect
and of its time-scale: depending on the THGEM geometry, the gain decreases or increases
and variations up to factors as large as 5 and even more have been observed;
stable conditions are reached after biasing the multipliers over days.  This
feature can be explained in terms of the charge mobility inside the dielectric,
which modifies the electric field.

\subsection{Gain time evolution observed with the Leopard setup}

\begin{figure}[h]
	\centering
	\includegraphics[width=.45\textwidth]{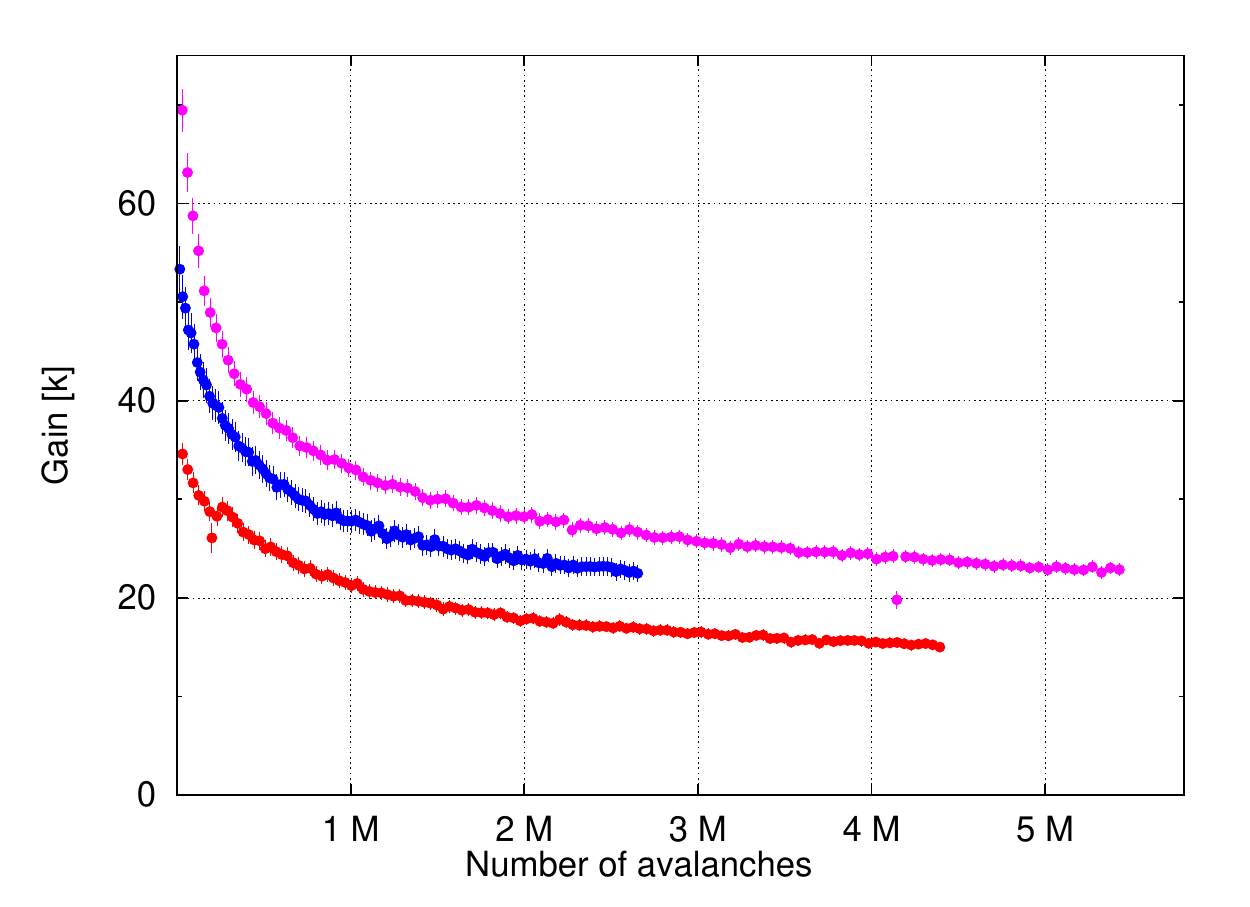}
	\caption{Charge-up curves for three different single hole in THGEM DESTRO-I; a statistic of 
                 one million events is collected in approximately 5 minutes.}
	\label{FigChargeUpSingleHoles}
	\end{figure}

Charging up processes have been observed during the performed Leopard measurements,
in particular the focus light-source allows to measure the single hole charging-up, 
as illustrated in Fig.\ref{FigChargeUpSingleHoles}.
\par 
When a THGEM is biased for long enough time, a second phenomenon is observed:
the yield increases and this effect is stronger in the large-rim THGEM C3HR-II.
A tentative explanation is the stronger electric field 
due to the motion of charges inside the dielectric, which reinforces the electric
field when a large rim is present and thus favours the photoelectron extraction.
The systematic studies to explore the yield variation are not part of the present paper.

\par
The gain variation by charging-up represents a systematic effect that affect the measurements. 
This effect has been partially reduced by two strategies:
biasing the THGEM for several hours before long-lasting measurements
and illuminating the area to be measured with unfocused UV LED light
before starting a measurement.
The residual systematic error on G related to the charging-up effect
is estimated by repeated measurements to be at the 5\% level.

\section{Gain studies}
\label{gain}

\subsection{Gain uniformity by comparing single hole gain}
\label{gain-uniformity}

\begin{figure}[h]
	\centering
	\includegraphics[width=.45\textwidth]{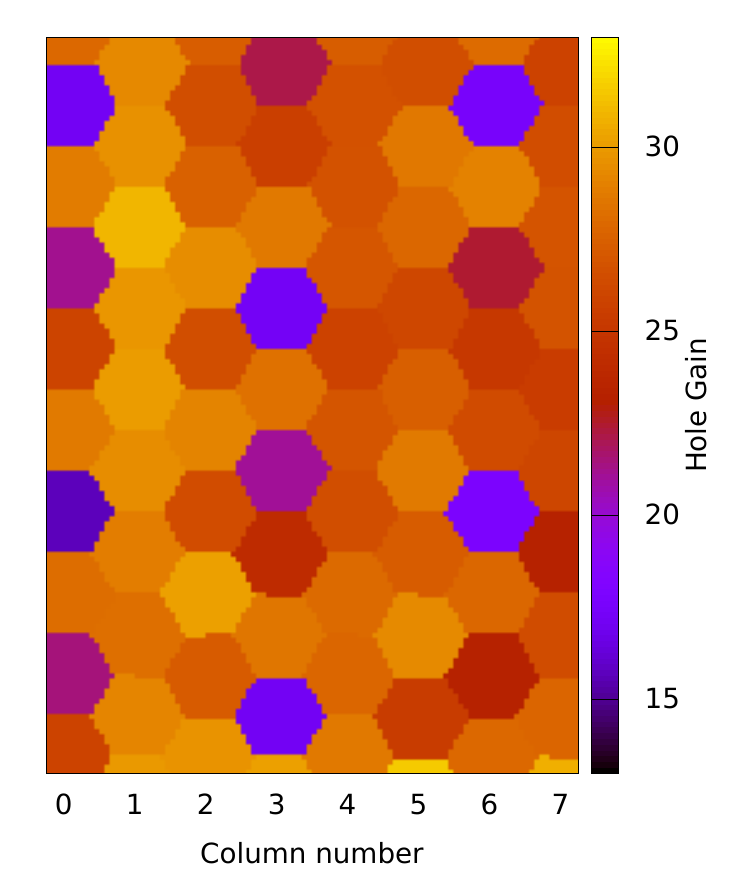}
	\caption{Map of hole-gain for a portion of the 
	DESTRO-I THGEM sample. The hexagonal regions correspond 
	to the hole-areas. The columns of hexagonal areas are labelled 
	with numbers.
		}
	\label{FigGainMap}
	\end{figure}

\begin{figure}[h]
	\centering
	\includegraphics[width=.45\textwidth]{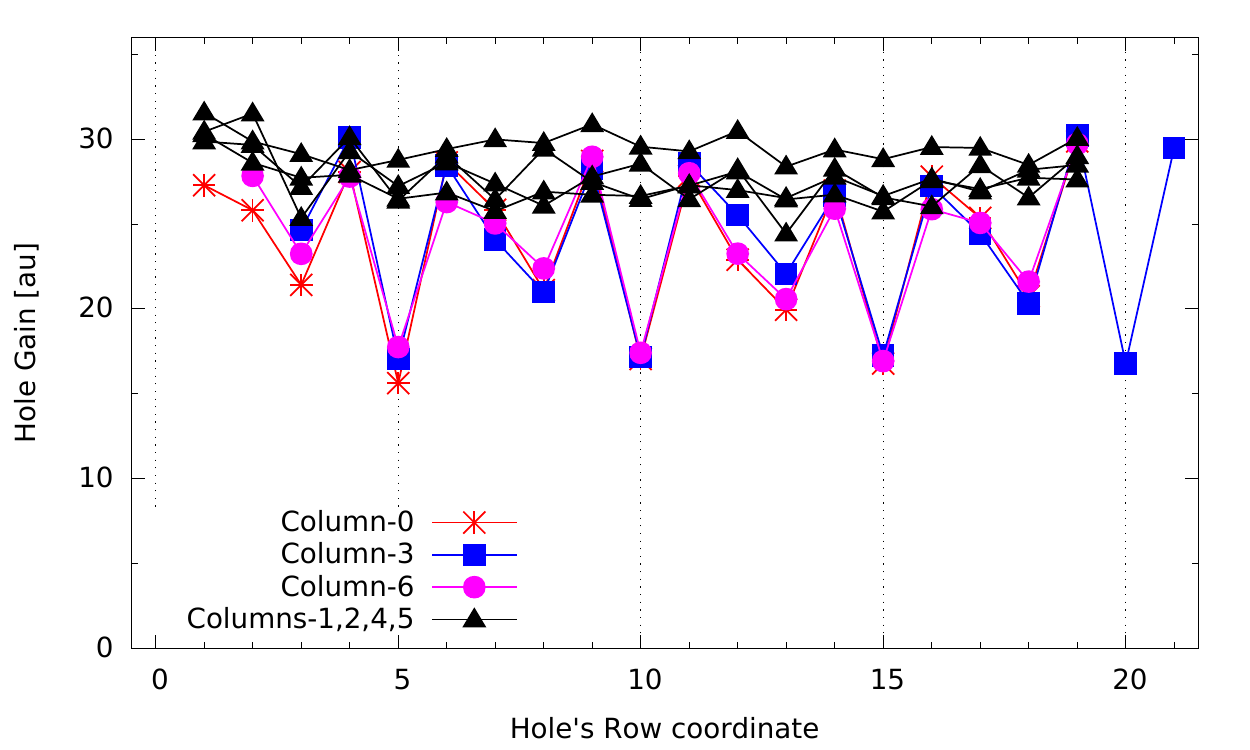}
	\caption{Hole-gain for the different hole-areas of the seven columns indicated in  Fig.~\ref{FigGainMap}.
		}
	\label{FigGainPilareffect}
	\end{figure}

\begin{figure}[h]
	\centering
	\includegraphics[width=.45\textwidth]{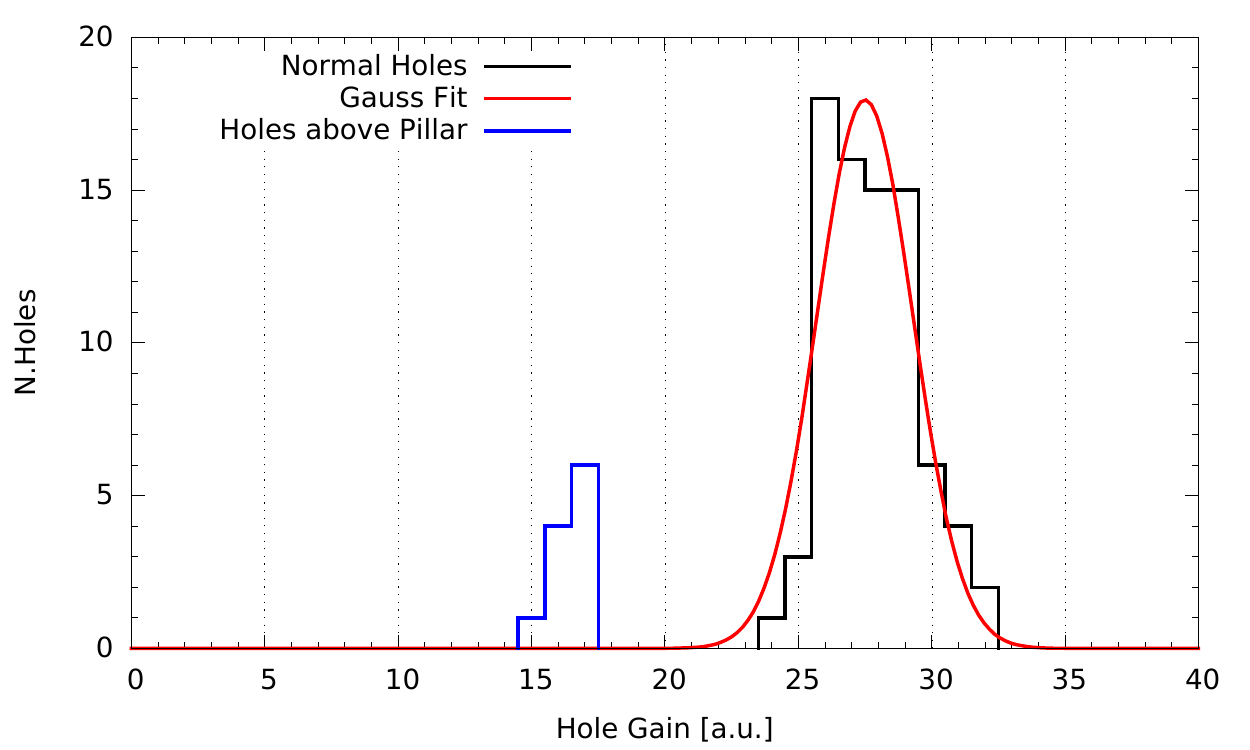}
	\caption{Hole-gain distribution for the DESTRO-I THGEM sample;
	black line: holes in the columns "1", "2", "4" and "5"; red line: gaussian fit of the distribution; blue line: holes fully aligned with MM pillars.	}
	\label{FigGainDistribution}
	\end{figure}

Figure~\ref{FigGainMap} presents the 2-D gain-map 
for a portion of the DESTRO-I THGEM sample. 
Each point at which the measurement is performed is
associated to the nearest hole; 
in fact, in the large majority of the cases, the detected photoelectron is guided by the 
the electric field to the nearest hole. The resulting regions associated to the holes are hexagons. 
The mean value of the gain measured in each hexagonal region is reported in the gain map.
The gain-value is dictated by the gain in the hole and by the multiplication in 
the MM portion below the hole itself. 
A marked pattern of different gain-values is observed:
in Fig.~\ref{FigGainMap} the columns of hexagonal areas that are labelled 
with numbers "0", "3" and "6" exhibit lower gain holes, which appear with regular periodicity, 
while the gain is almost constant in the other columns (Fig.~\ref{FigGainPilareffect}).
There is full correlation between the lower gain holes and the location of the 
pillars supporting the micromesh in the MM, which are dead areas of the MM itself. 
The distribution of the hole-gain for the columns "1", "2", "4" and "5" is shown in
Fig.~\ref{FigGainDistribution}; the distribution r.m.s. as provided by the gaussian fit is 
6.5\%, indicating an extremely good uniformity of the gain from this hole-by-hole study.
A second distribution is plotted showing the gain of holes fully aligned with the pillars: 
the resulting gain is approximately 60\% of that of holes far from the pillars.

\subsection{Gain uniformity versus radial distance from the hole centre}
\label{radial-gain}

Electric field calculation demonstrates that inside a 
hole of the THGEM the electric field is not constant radially. Therefore,
the average gain for an avalanche produced by  a single electron
depends on its radial position at the hole entrance.
Moreover, for the photoelectrons emitted from the converter at the THGEM surface, 
the same electrostatic calculations suggest that, 
the further the photoelectron is produced, the closer to the center of the hole it goes.
The diffusion of electrons along the drift path and in the avalanche process
smears the effect. A further element of complexity is 
the realistic calculation of the 
electric field in presence of charging up effects and 
charge displacement in the dielectric
material. An experimental investigation of the gain variation
versus the radial distance from the hole centre 
of the point where the photoelectron has been 
generated can be addressed with the Leopard setup
within the space resolution provided by the size of the focus light-spot.
\par
\begin{figure}[h]
	\centering
	\includegraphics[width=.45\textwidth]{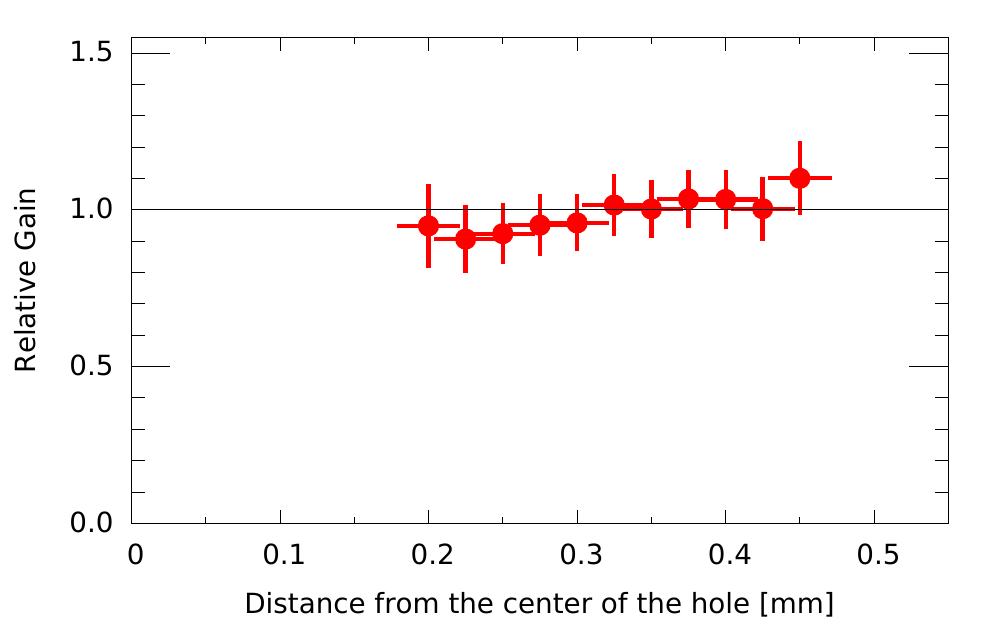}
	\caption{Relative gain versus the radial distance of 
                 the photoelectron emission point from the hole centre for THGEM DESTRO-I.}
	\label{FigGainR}
	\end{figure}

Dedicated fast scans have been performed on previously illuminated small areas,
thus avoiding the systematic effect due to charging-up and -down of the studied surface.
For each of the holes, the gain of all hole-area points, normalized to the 
hole-gain defined as the average gain of all the point of the hole-area, 
have been combined together: the result is shown in Fig.\ref{FigGainR}.
The measured distribution is flat within uncertainties, 
however does not exclude a moderate increase of the gain for the photoelectrons generated
far from the hole, which drift towards the center of the hole itself.

\section{Photoelectron extraction studies}
\label{extraction}

\subsection{The effective quantum efficiency in gaseous detectors}
\label{effective_qe}

\begin{figure}[h]
	\centering
	\includegraphics[width=.45\textwidth]{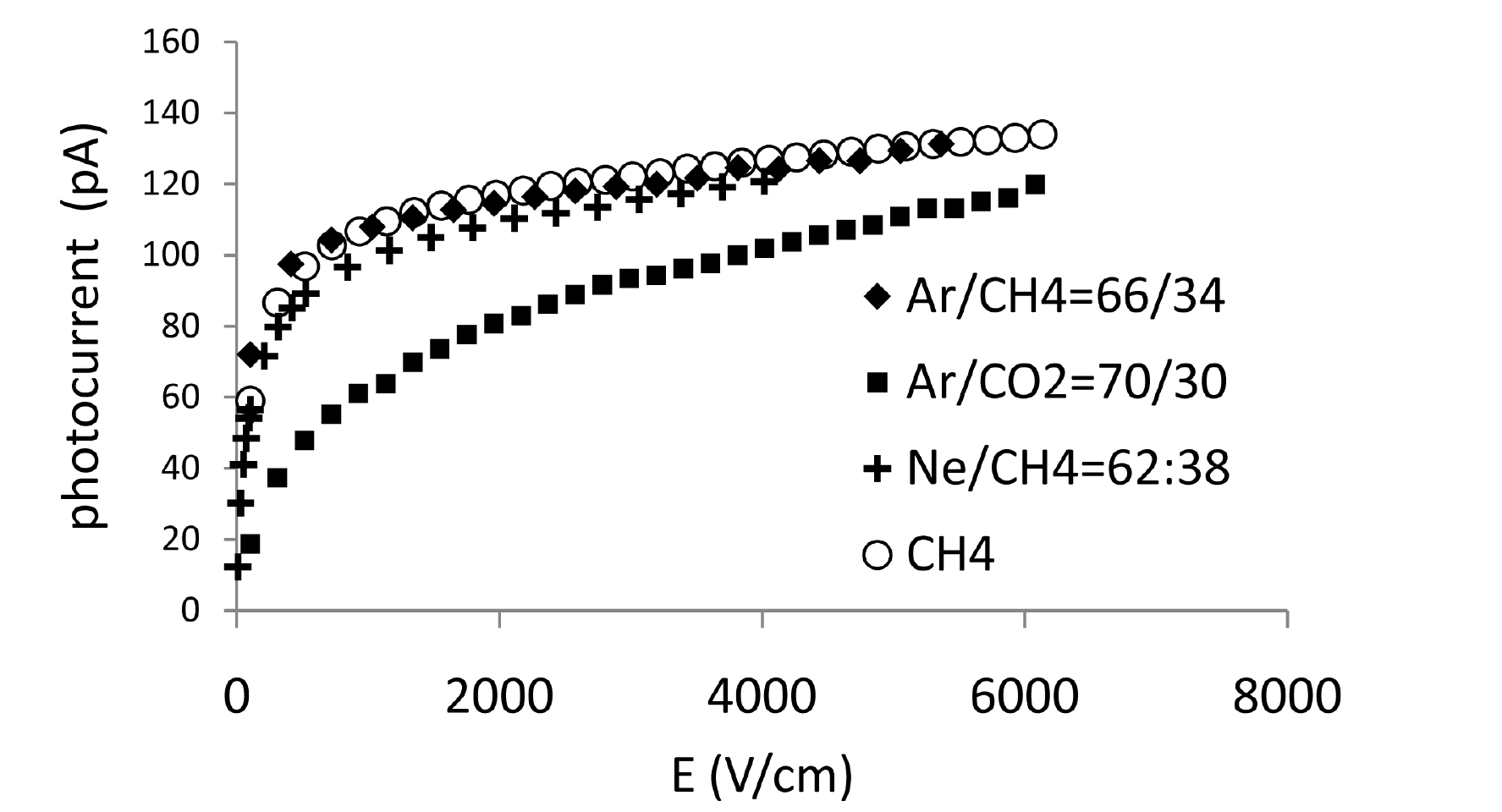}
	\caption{CsI photocurrent versus applied electric field in various
                 gases and gas mixtures at atmospheric pressure; systematic uncertainty
                 at the 1\% level\cite{noi_2010_qe}.}
	\label{FigOurQE}
	\end{figure}

The effective QE obtained in a gaseous detector depends on the gas used and the electric field at
the  photocathode surface. When the photoelectron elastically scatters  
off a gas molecule, the back-scattering probability is high and part of 
the photoelectrons impinge back on the photocathode where they are absorbed. 
Therefore, the effective QE is reduced. 
The characteristics of the gas molecules and the value of the
electric field accelerating the photoelectrons determine the
elastic scattering rate and thus the effective QE. 
Both parameters have been studied in the context
of RD26\cite{Breskin_1995_qe, DiMauro_1996_qe} 
and newly explored recently\cite{noi_2010_qe, Azevedo_2010_qe}. 
The response in different gas atmospheres has also been reproduced
by simulations\cite{DiMauro_1996_qe,Escala_2010_qe}. An example of the measurements 
performed is shown in Fig.~\ref{FigOurQE}. At atmospheric
pressure, the effective quantum efficiency increases very
steeply up to electric field values of about 1000 V/cm. 
At higher field values, the increase rate versus field is reduced,
even if it remains non-negligible. The highest quantum efficiency is obtained
in pure methane or in methane-argon mixtures, 
provided that the methane fraction is high ($>$~40\%). 

\par
A major question is the possibility to have good effective 
quantum efficiency on the whole reflective photocathode surface of a THGEM device. 
The field at the photocathode surface is the combination of the dipole field 
due to the voltage applied between the two THGEM faces and the external additional field, 
applied between the CsI coated THGEM face and an electrode placed above this surface, 
usually referred to as drift field in the literature.
The contribution of the dipole field results in a field pointing towards the photocathode surface, 
with variable intensity and orientation at different surface points. 
\par
The drift field is called direct when pointing outward from the THGEM surface, 
this being the direction used in tracking applications, 
and reversed when pointing toward the THGEM top surface. 
In this article positive values are assigned to the direct field 
and negative values to the reversed field.
When a reverse drift field is applied, part of the photoelectrons 
are collected at the electrode above the photocathode and they do not enter
the multiplication chain: they are lost. 
The direct drift field contributes in guiding the electrons towards the amplification holes. 
In this configuration the electric field accelerating the extracted photoelectons 
is the combination of the dipole field and the drift field. 
Therefore, the optimal drift field value is the one providing 
at the same time, the effective guidance to the electrons toward the holes 
and sufficient photoelectron acceleration.
\par
\begin{figure}[h]
	\centering
	\includegraphics[width=.45\textwidth]{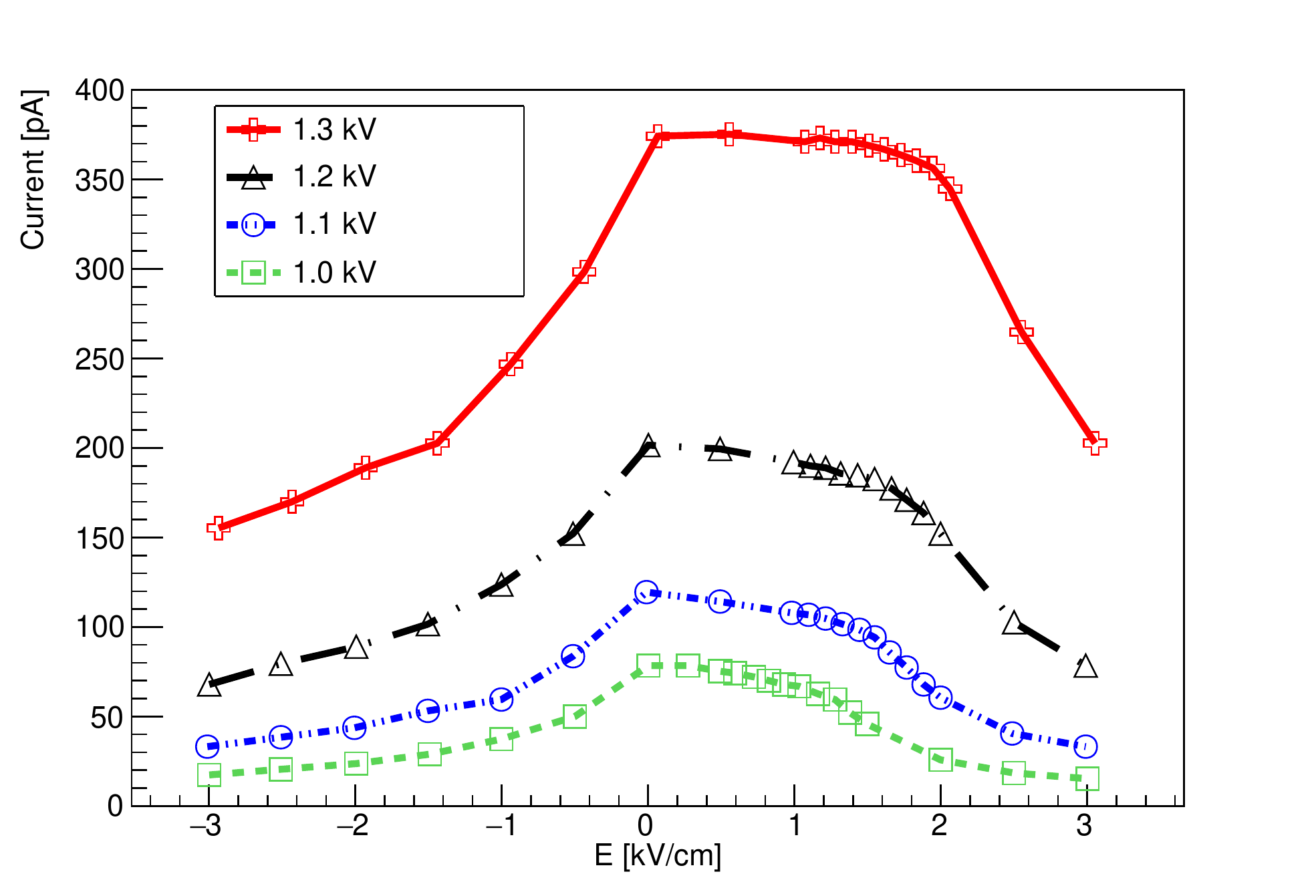}
	\caption{Anode current measured in a single THGEM detector
                 with CsI reflective photocathode versus the additional electric field
                 applied; the different point sets have been obtained for different values
                 of the potential $\Delta$V applied between the THGEM faces\cite{noi_2010_qe}.}
	\label{FigElenaCups}
	\end{figure}

A first optimization study has been performed measuring photocurrents\cite{noi_2010_qe}. 
A single THGEM layer with CsI coating illuminated with a UV lamp has been used 
and the current at the detector anode has been measured.
The plots in Fig.~\ref{FigElenaCups} clearly indicate a
sharp current decrease for reversed drift field, as expected,
and a rough plateau for moderate values of the direct drift field, 
followed by a drop when the total field pushes the photoelectrons back to the photocathode. 
Moreover, the current drops at lower values of the direct drift field
when the dipole field is lower, as intuitively expected.
This observation confirms the relevance of a high dipole field.
These first measurements, even if confirming the global picture,
suffer of a systematic limitation: the measured currents
depends both on the the number of extracted photoelectrons 
and the gain, which is not constant varying the drift field.

\par
Electrostatic calculations also indicate that 
higher dipole field at the THGEM surface can be obtained increasing
the ratio R of the hole diameter to the hole pitch.
At the same time, when R is large, the fraction of the THGEM surface that can be
coated is reduced; for R~=~0.5, this fraction is 77\%. The two
competing requests dictate a strong constrain on the ratio
value and suggest to adopt geometries with R=0.5.

\subsection{Direct observation of the effective quantum efficiency using THGEM as photocathode substrate}

Images obtained with the Leopard setup provide
a deeper understanding of the role of the drift field.
When the drift field is optimized, photoelectrons from the whole 
THGEM surface are effectively extracted and multiplied.
The relevant plots are 2-D $Y_{extr}$-maps.
The studies illustrated in this subsection
have been performed using the THGEM M1-III.
The same  number of events has been collected at 
each point, thus making the results within a map and those 
related to maps obtained in different conditions 
directly comparable.  
\par
\begin{figure}[h]
	\centering
	\includegraphics[width=.45\textwidth]{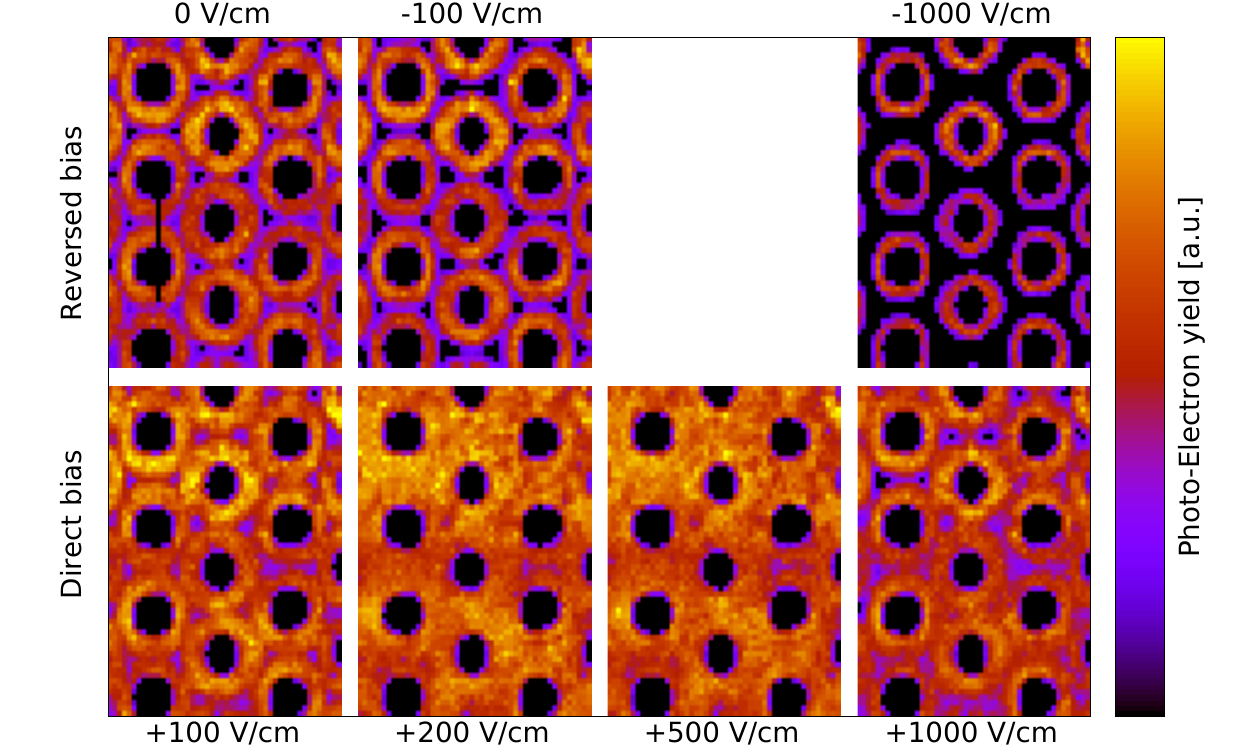}
    \caption{Y$_{extr}$-maps for fixed THGEM bias voltage varying
    the external drift field applied in front of the photocathode.}
	\label{FigDriftFieldScan_Plot201906}
	\end{figure}

Figure~\ref{FigDriftFieldScan_Plot201906} presents a the $Y_{extr}$-maps 
obtained for fixed THGEM bias voltage varying the drift field. 
The maps indicate as optimal range for the drift field
the range 200-500 V/cm of the direct field.
\par
\begin{figure}[h]
	\centering
	\includegraphics[width=.45\textwidth]{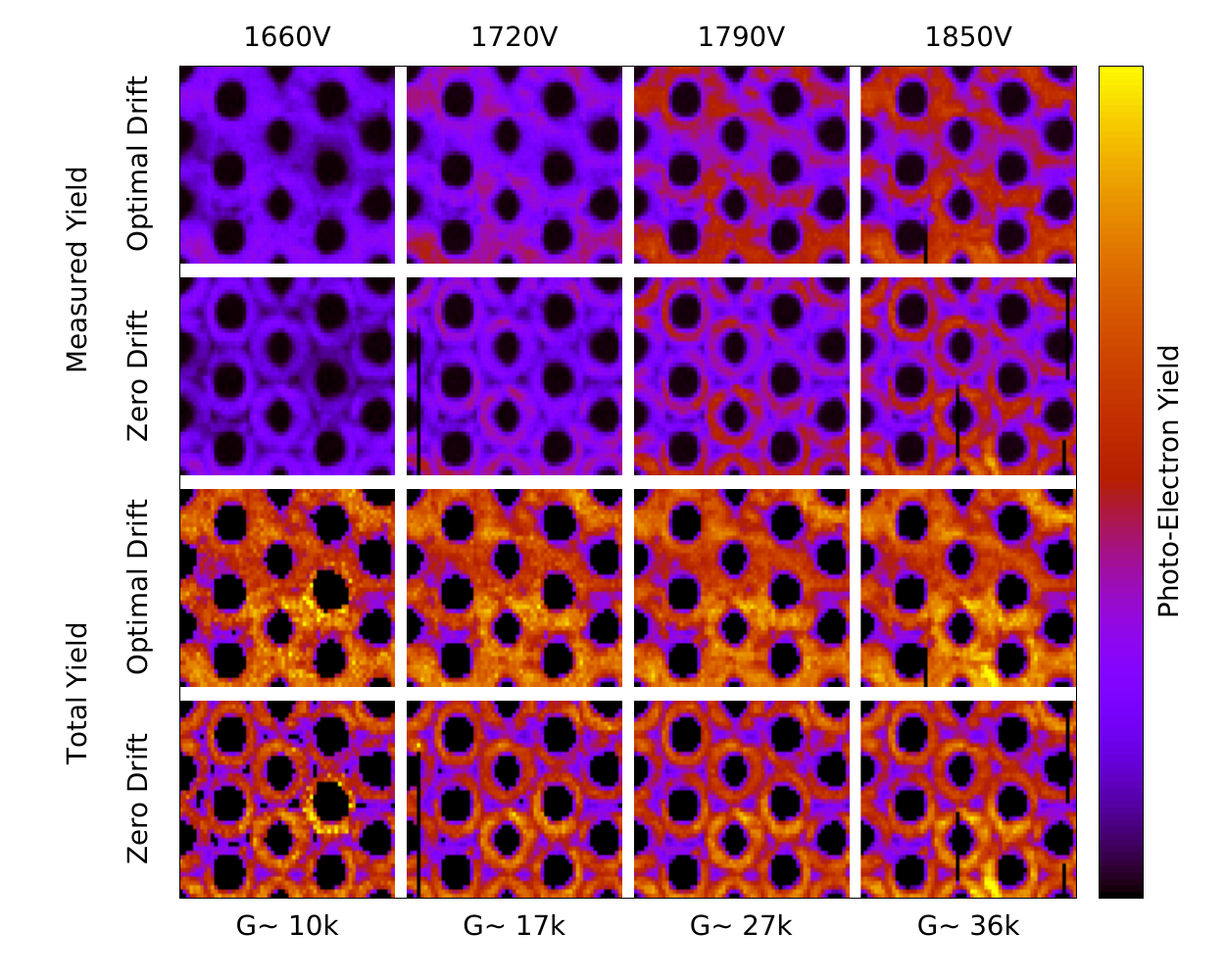}
	\caption{The Y-maps and the $Y_{extr}$-maps obtained applying two fixed value 
	of the drift field (0, optimal drift field: 200 V/cm) and  varying the THGEM bias voltage, 
	which results in an important variation of the detector gain, as indicated. }
	\label{FigDeltaVAndDriftScanMaps}
	\end{figure}

The Y-maps and $Y_{extr}$-maps obtained for two fixed 
value of the drift field varying the THGEM bias voltage
are shown in Fig.~\ref{FigDeltaVAndDriftScanMaps}. 
The typical noise value during these measurements
is around 2300 electrons equivalent. The corresponding 5-$\sigma$ threshold value is 11500
electrons equivalent. The resulting detection
efficiency  is about 30\% at a gain of 10k gain obtained at 
1660~V and larger than 72\% at a gain of 36k obtained at 1850~V.
The $Y_{extr}$-maps show that, in spite of the 
relevant variation of the detector total gain, 
no relevant variation of the effective quantum efficiency is observed 
when the optimized drift field is applied.
At the same time, the Y-maps underlay the relevance 
of operating at high detector gain when the noise level imposes high 
threshold values of the read-out front-end electronics.

\par
\begin{figure}[h]
	\centering
 	\includegraphics[width=.45\textwidth]{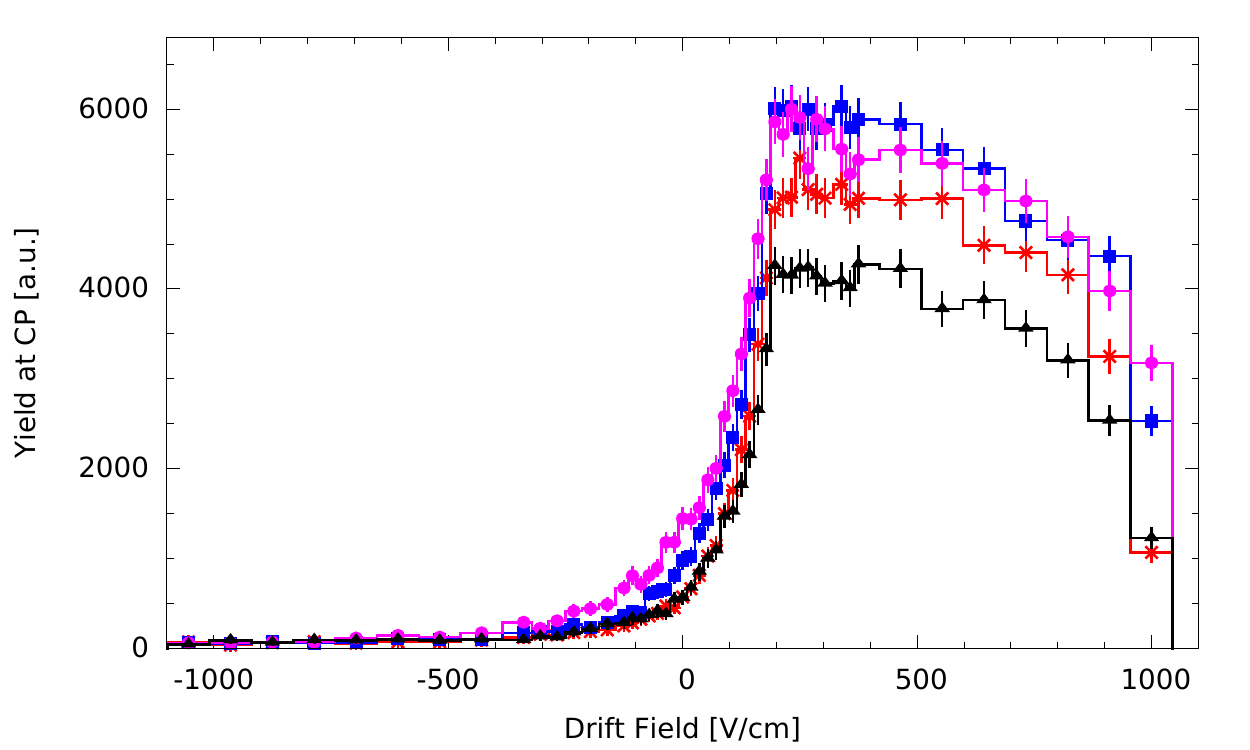}
	\caption{The photoelectron yield at four different CPs versus the  
                applied drift field.}
	\label{fig:CriticalPointScan}
    \end{figure}

The electric field configuration at the THGEM surface is particularly low
at the points which are at equal distance from the three nearest holes
resulted from simple symmetry considerations.
Correspondingly, it is expected that the photoelectron extraction is more problematic 
at this Critical Points (CP). 
The expectation is confirmed by the images in Fig.~\ref{FigDeltaVAndDriftScanMaps}, where 
it can also be observed that, for optimized values of the drift field, 
the inefficiency at the CPs is overcome.
The rapid evolution of the photoelectron extraction efficiency
at the CP is illustrated in Fig.~\ref{fig:CriticalPointScan}.

\par
\begin{figure}[h]
	\centering
	\includegraphics[width=.45\textwidth]{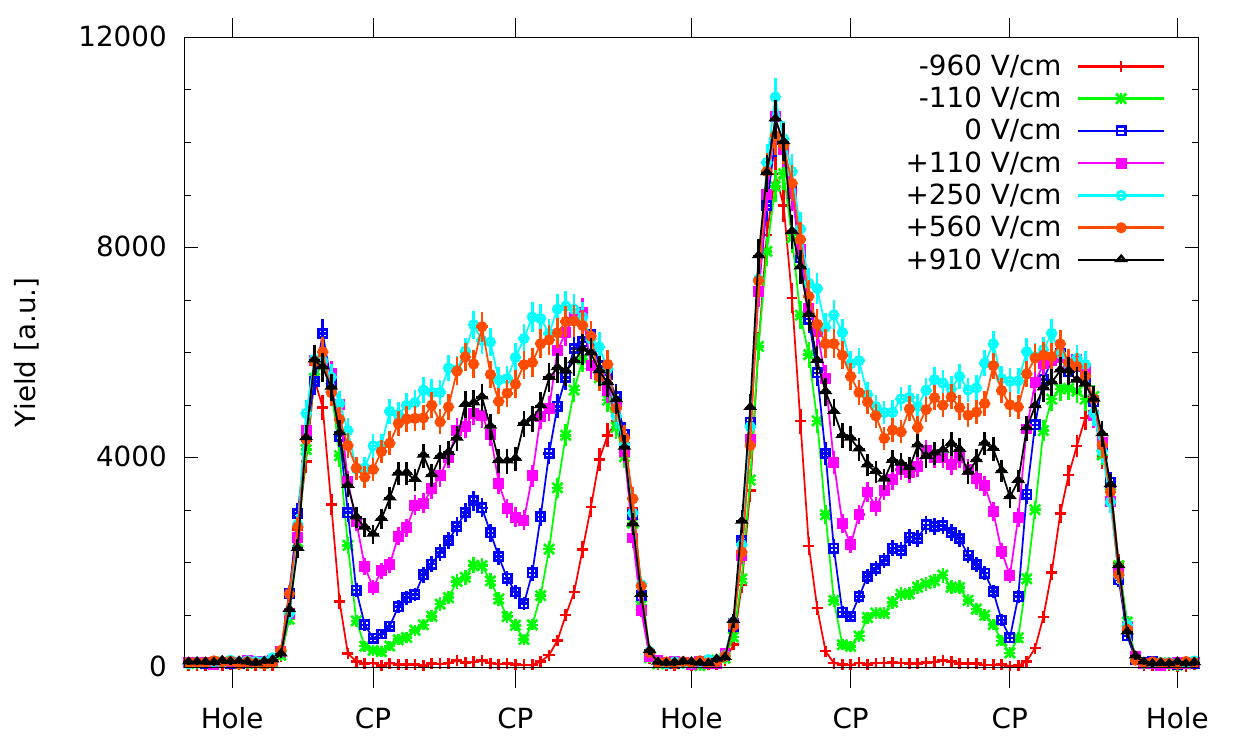}
	\caption{Y$_{extr}$ from the one-dimensional scan along a critical line portion, 
    including three holes and four CPs, for seven different values of the
    drift fields. }
	\label{fig:CriticalLineSamples}
	\end{figure}
\begin{figure}[h!]
	\centering
	\includegraphics[width=.45\textwidth]{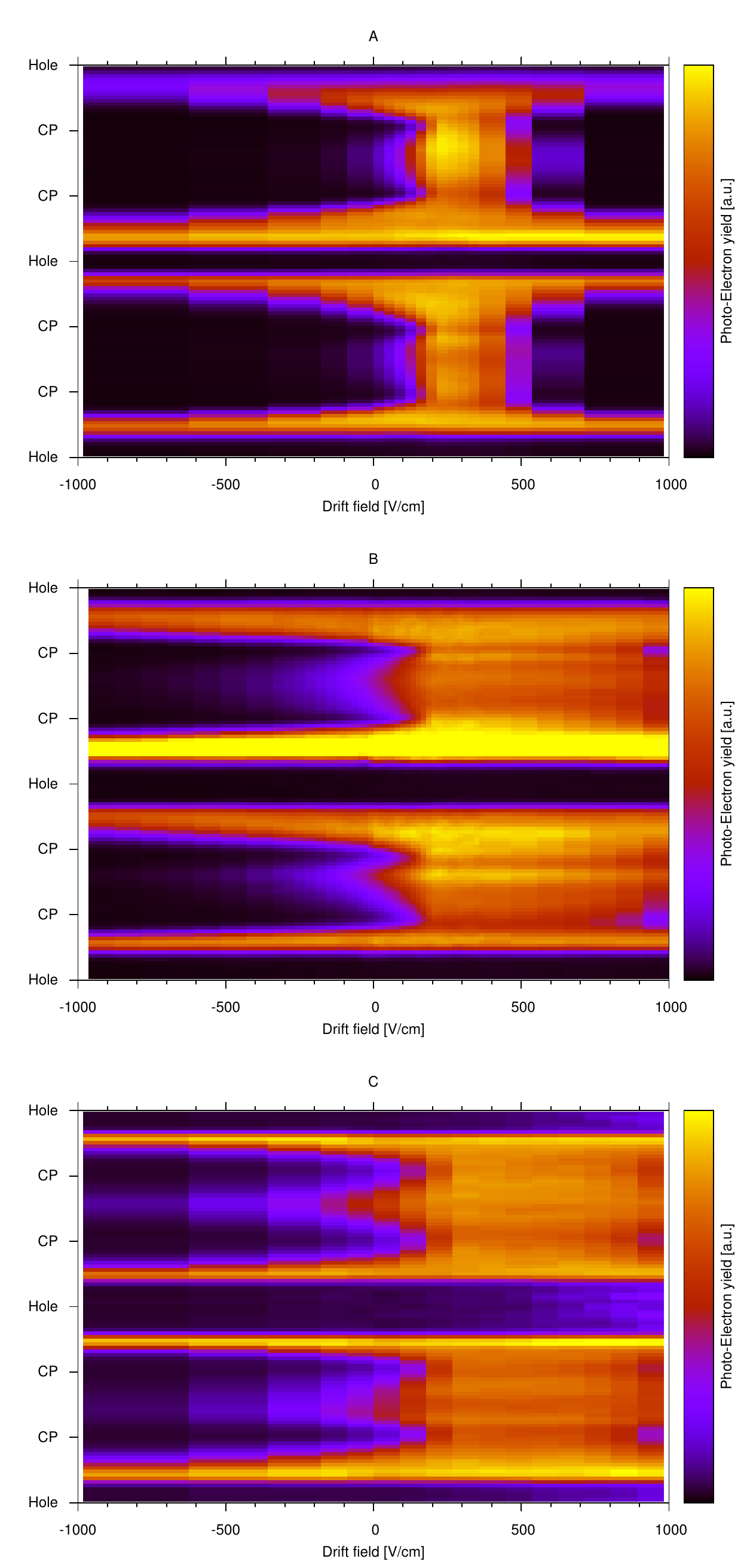}
	\caption{Y$_{extr}$ from the one-dimensional scans along a critical line (y-axis) 
	varying the drift field (x-axis). Three different THGEMs have been 
	used: M2.1-II (A), M1-III (B), M2.4-G (C).}
	\label{fig:CLS_Multiplot_abc}
	\end{figure}

The photoelectron extraction efficiency can also be studied 
with a quicker approach which does not require two-dimensional scans: 
one-dimensional scans along lines connecting a row of CPs, 
critical line, are performed.
Figure~\ref{fig:CriticalLineSamples} presents a set of 
seven one-dimensional scans obtained
by varying the drift field;
where the
the critical line crosses three holes and four CPs.
The results of scans along the same critical line varying the drift field are summarized in Fig.~\ref{fig:CLS_Multiplot_abc}, B:
the region of optimal drift field and the 
photoelectron yield evolution
along the critical line are clearly
visible.

\subsection{Comparison of the photoelectron extraction in  THGEMs with different geometry.}
\label{different-thgems}

One-dimensional scans along a  critical line varying the applied drift field have been performed for
THGEMs with different geometrical parameters.
The results are presented in Fig.s~\ref{fig:CLS_Multiplot_abc}, 
A for THGEMs M2.4-G and C for THGEM
M2.1-II. These plots have to be compared with the one presented 
in Fig.~\ref{fig:CLS_Multiplot_abc}, B obtained using THGEM M1-III. 
\par
THGEM M2.4-G is thicker than THGEM M1-III; 
in a thicker THGEM, a larger fraction of the 
dipole electric field is confined in the holes 
and the high field region, where the multiplication 
takes place, is longer. Therefore, larger gain can be obtained. 
At the same time, it can be expected that the photoelectron
extraction at the CPs is more marginal 
because the dipole electric field is more feeble there. 
This aspect is not confirmed by our measurement: THGEM M2.4-G
geometrical parameters are as valid as the ones of THGEM M1-III.
\par
In THGEM M2.1-II,  the hole diameter is smaller than in THGEM 1-III;
also in this case, a larger fraction of the dipole electric 
field is confined in the holes and expectations are similar 
to those of the previous case: higher gain and more difficult
photoelectron extraction at the CPs. The plots indicate 
that complete photoelectron extraction is possible also for this
geometry, even if  the range of optimal drift field 
values is more limited. 
This indication is of particular interest because, 
thanks to the reduced R value of THGEM M2.1-II (R= 0.375) 
the surface available for CsI coating is larger: 87\% to be 
compared with the value of 77\% for THGEM M2.1-III.
Therefore, THGEMs with M2.1-II  geometrical parameters can be considered as
photocathode substrates in MPGD-based photon detectors.

\section{Photoelectron extraction from CsI and from gold in gaseous atmosphere}
\label{csi-gold}

The measurements described in Sec.~\ref{extraction} 
have been performed extracting photoelectrons from 
gold-coated surfaces. The 
gaseous photon detectors are operated with CsI photoconverters.
Therefore, we have performed a set of measurements to establish the 
portability of the results described in the previous sections to the
gaseous photon detectors as operated in experiments.
For these purposes we have compered measurements performed 
extracting photons from gold and from CsI using argon-methane gas mixtures.
\par
\begin{figure}[h]
	\centering
	\includegraphics[width=.45\textwidth]{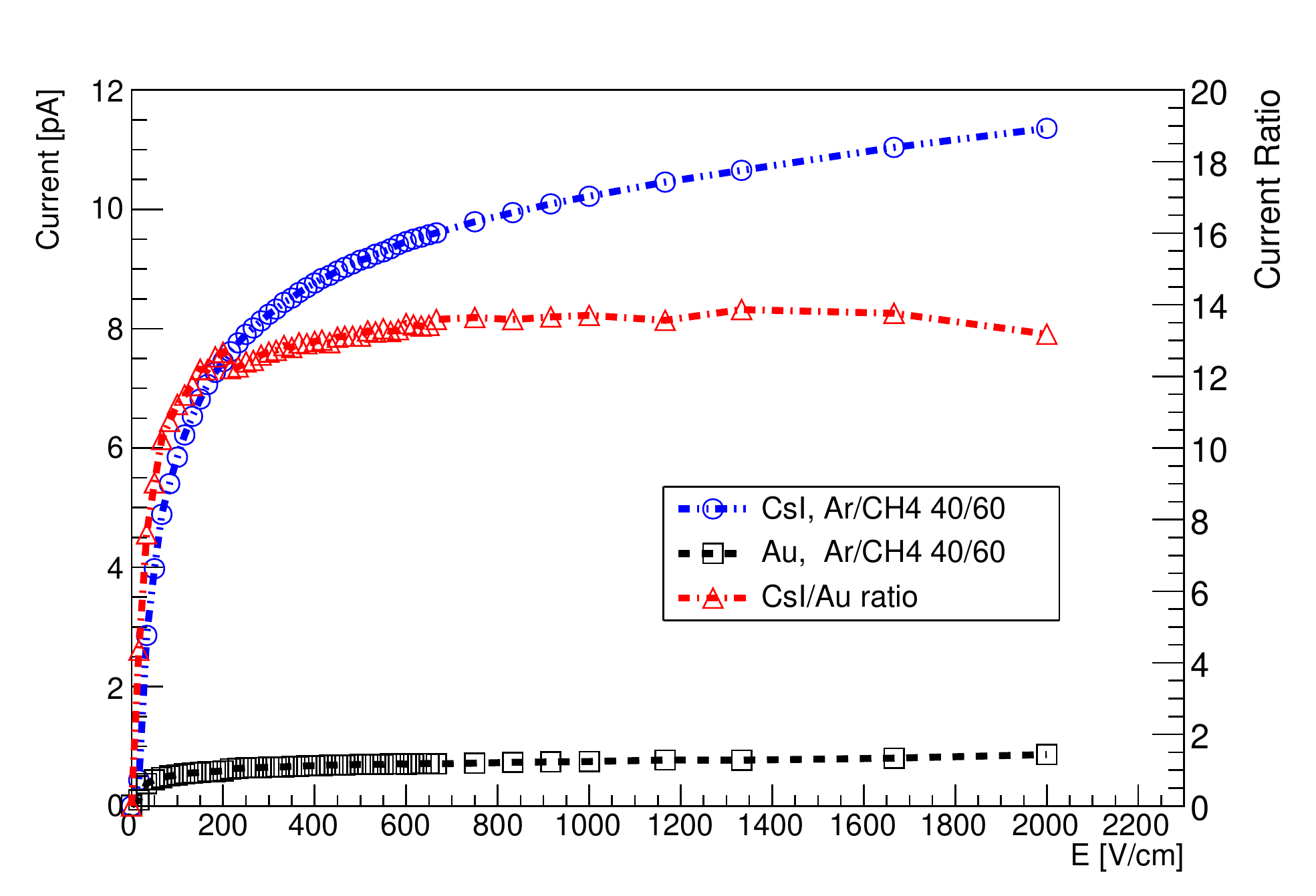}
	\caption{Current measured at the anode while illuminating 
    the CsI coated cathode (circles) and the gold coated cathode 
    (squares) versus the electric field in Ar:CH$_4$~=~60:40 atmosphere; 
    the ratio of the two currents is also shown (triangles). }
	\label{fig:40-60}
	\end{figure}

Two parallel plates are housed in a chamber filled with the
appropriate gas mixture and a voltage bias between them is applied. 
The plate acting as photocathode is
by fiberglass and it is coated with a layer of 35~$\mu$m of copper and superimposed layers of 
nickel (5~$\mu$m) and gold (0.5~$\mu$m); one more layer of CsI, 300~nm thick,
has been added to the CsI-photocathode.
The light from a deuterium lamp\footnote{Oriel type 63162} is guided via a quartz fibre
to the photocathode and the light impinges on it at 45$^o$.
The current generated by the extracted photoelectrons is measured 
with a picoampermeter\footnote{Keithley type 6485}. 
Results are presented in Fig.~\ref{fig:40-60} for  Ar:CH$_4$=60:40. 
The ratio of the anode current versus the electric field measured 
using the CsI coated cathode and the gold coated one exhibits a very modest increase 
for increasing values of the electric field.
This behavior indicates that the results concerning the photoelectron
extraction from THGEM photocathodes obtained with the gold coated 
THGEMs can be extended to the THGEMs with CsI coating, 
even if the dependence of the photoelectron extraction on the electric field 
is slightly more marked when CsI is present.

\section{Conclusions}
\label{conclusions}
Direct measurements of the properties of THGEM reflective photocathodes
have been performed scanning THGEM surfaces by a high resolution optical system 
and detecting photoelectrons. 
\par
The results of this laboratory investigation concern 
the THGEM gain uniformity in hole-by-hole measurements 
and versus the distance from the 
nearest hole, photoelectron extraction 
with particular care dedicated to the extraction location,
the biasing voltage applied to the THGEM,
the electric field above the photocathode surface
and the extraction properties of THGEMs with different geometry.
The direct measurement confirm the indication provided by 
precedent indirect measurements.
\par
Photoelectron extraction from CsI-coated and from gold-coated 
photocathodes in gaseous atmosphere
have been compared. It is so possible to extend the results obtained for gold-coated THGEMs 
to CsI coated devices.
\par 
The results have provided precious information to guide the design, 
construction and operation of 
gaseous photon detectors where THGEMs are used as photocathode substrates.


\section*{Acknowledgments}
 
This work has been performed in the framework of the
RD51 collaboration and the COMPASS RICH-1 upgrade;
the authors are grateful to the Colleagues 
of both collaborations
for constant support and encouragement.
\par 
The activity is supported
in part by the European Community, 
within the H2020 project AIDA-2020, GA no. 654168.


\bibliographystyle{elsarticle-num}

\end{document}